\documentclass[review]{elsarticle}
\biboptions{sort&compress} 

\usepackage{epsfig,amsmath,amssymb}%,hyperref}
\usepackage{enumerate}
\usepackage{subfigure}
\usepackage{graphicx} 
\usepackage{epstopdf}
\usepackage{booktabs}
\usepackage{multirow}
\usepackage{xcolor}
\usepackage{threeparttable}
\journal{arXiv}

\usepackage{amsmath,amssymb}
\usepackage{subfigure,algorithm,algorithmic}

\makeatletter
\newenvironment{breakablealgorithm}
{% \begin{breakablealgorithm}
	\begin{center}
		\refstepcounter{algorithm}% New algorithm
		\hrule height.8pt depth0pt \kern2pt% \@fs@pre for \@fs@ruled
		\renewcommand{\caption}[2][\relax]{% Make a new \caption
			{\raggedright\textbf{\ALG@name~\thealgorithm} ##2\par}%
			\ifx\relax##1\relax % #1 is \relax
			\addcontentsline{loa}{algorithm}{\protect\numberline{\thealgorithm}##2}%
			\else % #1 is not \relax
			\addcontentsline{loa}{algorithm}{\protect\numberline{\thealgorithm}##1}%
			\fi
			\kern2pt\hrule\kern2pt
		}
	}{% \end{breakablealgorithm}
		\kern2pt\hrule\relax% \@fs@post for \@fs@ruled
	\end{center}
}
\makeatother

\usepackage{amsmath}

        \makeatletter
        \def\fps@eqnfloat{!t}
        \def\ftype@eqnfloat{4}
        
        \newenvironment{eqnfloat*}
               {\@dblfloat{eqnfloat}}
               {\end@dblfloat}
        \makeatother
\makeatletter
\let\oldabs\abs
\def\abs{\@ifstar{\oldabs}{\oldabs*}}
\let\oldnorm\norm
\def\norm{\@ifstar{\oldnorm}{\oldnorm*}}
\makeatother

\renewenvironment{thebibliography}[1]{%
	\begin{oldthebibliography}{#1}%
		\setlength{\parskip}{0ex}%
		\setlength{\itemsep}{0ex}%
	}%
	{%
	\end{oldthebibliography}%
}%

\begin{document}
\begin{frontmatter}
% Title.
% ------
\title{Joint Design of the Transmit and Receive Weights for Coherent FDA Radar \tnoteref{t1}}
\tnotetext[t1]{This work was supported by National Natural Science Foundation of China under grant 62171092.}

%\author{Wenkai Jia}
%\ead{mrwenkaij@126.com}
%
%\author{Wen-Qin Wang\corref{cor1}}
%\ead{wqwang@uestc.edu.cn}
%
%\author{Shungsheng Zhang}
%\ead{zhangss@uestc.edu.cn}
%
%\cortext[cor1]{Corresponding author.}
%
%
%
%\address{School of Information and Communication Engineering, University of Electronic Science and Technology of China, 611731, Chengdu, P. R. China.}

\author[addr]{Wenkai Jia\corref{cor}}
\address[addr]{School of Information and Communication Engineering, University of Electronic Science and Technology of China, 611731, Chengdu, P. R. China.}
\ead{mrwenkaij@126.com}

\author[addr]{Wen-Qin Wang}
\cortext[cor]{Corresponding author.}
\ead{wqwang@uestc.edu.cn}

\author[addrr]{Shunsheng Zhang}
\address[addrr]{Research Institute of Electronic Science and Technology, University of Electronic Science and Technology of China, 611731, Chengdu, P. R. China.}
\ead{zhangss@uestc.edu.cn}

%% Information to appear on the title page:
%
%\ititle{Low-Complexity Uncertainty Set Based Robust Adaptive Beamforming for Passive Sonar}
%\iauthor{\quad\quad\quad Samuel~D.~Somasundaram, Naveed~R.~Butt, \newline Andreas~Jakobsson, and Les~Hart}
%\idate{2015}
%\ipublishedin{IEEE  Journal of Oceanic Engineering}
%\idoi{10.1109/JOE.2015.2474495}
%\iaddress{Mathematical Statistics\\
%  Centre for Mathematical Sciences\\
%  Lund University}
%%\publishedIEEE{2015}
%
%\makeititle

\begin{abstract}
Frequency diverse array (FDA) differs from conventional array techniques in that it imposes an additional frequency offset (FO) across the array elements. The use of FO provides the FDA with the controllable degree of freedom in range dimension, offering preferable performance in joint angle and range localization, range-ambiguous clutter suppression, and low probability of intercept, as compared to its phased-array or multiple-input multiple-output (MIMO) counterparts. In particular, the FO of the coherent FDA is much smaller than the bandwidth of the baseband waveform, capable of obtaining higher transmit gain and output signal-to-interference-plus-noise ratio (SINR).
In this paper, we investigate the problem of joint design of the transmit and receive weights for coherent FDA radar systems.
The design problem is formulated as the maximization of the ratio of the power in the desired two-dimensional range-angle space to the power in the entire area, subject to an energy constraint that limits the emitted energy of each transmit antenna and a similarity constraint such that a good transmit beampattern can be guaranteed.
Due to the resultant problem is NP-hard, therefore, a sequential optimization method based on semidefinite relaxation (SDR) technique is developed.
Numerical simulations are provided to demonstrate the effectiveness of the proposed scheme.
\end{abstract}
\end{frontmatter}

\section{Introduction}
%\noindent {\em Notation:} In this paper, vectors (matrices) will be denoted by lower (upper) case boldface letters. The diag$(\x)$ operator takes the vector $\x$ and inserts it on the diagonal of a matrix with zeros elsewhere. Furthermore, $(\cdot)^T$ and $(\cdot)^*$ denotes the transpose and the complex conjugate transpose respectively, and tr$(\mathbf{X})$ denotes the trace of the matrix $\mathbf{X}$. We abbreviate with respect to as w.r.t and use SNR for short for signal to noise ratio.
%\\ \newline
\noindent
The architecture of multiple transmitters and receivers enables the multiple-input multiple-output (MIMO) radar \cite{li2008mimo} superiority to phased-array radar in parameter identifiability \cite{li2007parameter} and target detection \cite{xu2008target}. 
Distributed MIMO \cite{haimovich2007mimo} that employs widely separated transmit/receive antennas over a geographic area exploits multiple paths through which the signals propagate from the transmit antennas to the receive antennas can take advantage of the spatial properties of extended targets \cite{fishler2006spatial} to obtain a diversity gain for Doppler
processing \cite{wang2011moving} and high resolution target localization \cite{lehmann2006high}.
Colocated MIMO \cite{li2007mimo} that transmit/receive elements are closely spaced offers higher sensitivity to detecting slowly moving targets \cite{forsythe2004multiple} and direct applicability of adaptive array techniques \cite{xu2006adaptive,4194775}. Besides, MIMO radar with the advantage of transmit waveform diversity \cite{friedlander2007waveform,7762192} can transmit via its antennas multiple probing signals that may be much optimized to obtain the flexibility for transmit beampattern design \cite{stoica2008waveform,7435338}. 
Spectrum sensing-based MIMO radar designs have also attracted much attention \cite{7838312,8574991,9052442}.
However, its beam steering is fixed in an angle for all the ranges.
The range-independent directivity brings serious limitations to
many applications, such as joint range and angle estimation,
range ambiguous clutter suppression and non-desirable range-dependent interferences mitigation.
Unlike the MIMO radar with a single carrier frequency, frequency diverse array (FDA) adds small frequency offsets (FOs) across the array elements, and the resulting controllable degrees of freedom in the range dimension provides potentials for range-dependent energy management and supports many promising applications
\cite{antonik2006multi,wang2016overview,wang2018ultrawideband}. 

For three-dimensional target localization, \cite{xu2016adaptive} proposed an adaptive range-angle-Doppler processing approach by exploiting the increased degree of freedoms in transmit, receive, and pulse dimensions of FDA.
However, range-dependent FDA may suffer from ambiguities in the range-angle dimension during the target localization \cite{9440819}, and the time-variant transmission process of FDAs is also troublesome \cite{xu2015jjoint}.
To deal with these problems, the FDA-MIMO technique is proposed \cite{6404099}, which configures a FO larger than the bandwidth of the baseband waveform on the basis of the standard FDA.
This technique enables FDA-MIMO to separate the transmit waveforms in the receiver, obtain the time-independent steering vector, and fulfill multiple tasks that the conventional phased-array and MIMO radars cannot handle \cite{8954891,9212375,9161264}.
For example, \cite{8954891} proposed an FDA-MIMO transmit-receive beamforming scheme with accurately controlled null regions, implemented by assigning artificial interferences with prescribed powers within the given rectangular regions, and applied it to imaging fields to overcome the problem of interference.
In order to achieve an optimal beampattern that approximates the desired beampattern in the range-angle plane while minimizing the cross-correlation sidelobes, utilizing the Hamming window based non-uniform FOs along the FDA-MIMO array, a transmit beamspace matrix design method based on alternating direction method of multipliers (ADMM) was proposed in \cite{basit2021transmit}.
In \cite{gui2020adaptive}, an adaptive FDA-MIMO transmit power allocation scheme was designed for spectral interference avoidance through maximizing the output signal-to-interference-plus-noise ratio (SINR), which shows that FDA radar is able to suppress mainlobe spectral interferences.
Moreover, \cite{wang2021lpi} investigated the low probability of intercept (LPI) properties of FDA-MIMO transmitted signals under typical passive reconnaissance techniques.
Particularly, in \cite{wang2016cognitive}, cognitive FDA-MIMO with situational awareness was introduced to avoid undesired strong interferences.

However, similar to co-located MIMO that deploys orthogonal baseband waveforms, the waveforms transmitted by FDA-MIMO are orthogonal in the frequency domain, resulting in underutilized FDA transmission gain.
Since the FO used by coherent FDA is much smaller than FDA-MIMO, it not only has a controllable beampattern in the range dimension, but also provides a higher transmission gain.  But its reception has always been a research difficulty.
The multi-channel matched filtering structure proposed in \cite{gui2017coherent} is proved to be the most effective coherent FDA receiver.
Particularly, Xu \emph{et al}. proposed an enhanced transmit-receive beamforming approach to achieve the time-invariant and symmetrical beampattern with only a single maximum value in range-angle space \cite{xu2020enhanced}.
Using a similar approach, a range-angle-decoupled equivalent beampattern with low sidelobe in both range and angle domains is also presented in \cite{xu2018low}.
However, the problem is that they only designed the receive weights, but the transmit weights were not optimized. More seriously, the output performance will degrade significantly in some scenarios if uncontrolled transmit power allocation is adopted. An intuitive method is to jointly optimize the transmit and receive weights via power maximization strategy \cite{6656878}. 
Note that the joint transmit-receive design of the MIMO radar shows better performance than the conventional omnidirectional MIMO radar \cite{9693236,9865367}.
To our knowledge, joint design of transmit and receive weights for coherent FDA was not discussed in open literature.

In this paper, we address the problem of joint design of the transmit and receive weights for coherent FDA radar systems, aiming to maximize the received RPDE (ratio of the power in the desired two-dimensional range-angle space to the power in the entire area).
Meanwhile, to guarantee the optimal transmission characteristics, we derive a time-independent FDA transmit beampattern, taking into account a similarity constraint between the designed transmit weights and the reference weights.
Then, we formulate an optimization problem, which involves a nonconvex fractional quadratic objective and several quadratic constraints.
To deal with the resulting NP-hard problem, an iterative method is devised where each update is processed using the semidefinite relaxation (SDR) technique.
In the simulations, various numerical examples are provided to assess the performance of the proposed scheme.

The remaining sections are organized as follows.
In Section \ref{sec:a}, the coherent FDA and time-independent transmit beampattern signal models are presented.
In Section \ref{sec:b}, we discuss the design constraints and formulate a joint optimization problem for transmit and receive weights.
Next, the proposed algorithms based on SDR techniques are described in Section \ref{sec:c}.
Finally, Section \ref{sec:d} provides the simulation results and conclusions are drawn in Section \ref{sec:e}.

\emph{Notation:} Lower case letters ${\mathbf{a}}$ and upper case letters ${\mathbf{A}}$, respectively, represent vectors and matrices.
Conjugate, transpose, and conjugate transpose operators are denoted by the symbols ${\left(  \cdot  \right)^c}$, ${\left(  \cdot  \right)^T}$, and ${\left(  \cdot  \right)^H}$, respectively.
For a matrix ${\mathbf{A}}$, we use ${\mathbf{A}}\left( {i,j} \right)$ as the $\left( {i,j} \right)$-th element of ${\mathbf{A}}$, $\operatorname{Tr} \left\{ {\mathbf{A}} \right\}$ as the trace of ${\mathbf{A}}$ , $\operatorname{Rank} \left\{ {\mathbf{A}} \right\}$ as the rank of ${\mathbf{A}}$, and  ${\lambda _{\max ,vec}}\left\{ {\mathbf{A}} \right\}$ as the principal eigenvector of ${\mathbf{A}}$.
The symbol $\operatorname{diag} \left\{ {\mathbf{a}} \right\}$ denotes the diagonal matrix with the diagonal entries formed by ${\mathbf{a}}$ and $\left\| {\mathbf{a}} \right\|$ denotes the Euclidean norm of ${\mathbf{a}}$.
${{\mathbf{I}}_{M}}$ and ${{\mathbf{1}}_{M\times M}}$ represent the identity matrix and all-ones matrix with size $M \times M$, respectively.
We let $ \succeq$ stand for generalized matrix inequality and $*$ for convolution operator.
Hadamard matrix product and Kronecker matrix product are expressed as $ \odot $ and $ \otimes$, respectively.

\section{Signal model}\label{sec:a}
\noindent
Consider a standard uniform FDA radar system with $M$ co-located transmit antennas and $N$ co-located receive antennas \cite{antonik2006frequency}.
For the $m$-th transmit antenna, a beamforming weight $w_m$ is deployed to transmit a baseband waveform $s(t)$ with unit energy, i.e.,
\begin{equation}\label{eq:LASSO}
\int_{{{T}_{p}}}{s\left( t \right){{s}^{c}}}\left( t \right)\text{d}t=1.
\end{equation}
where $T_p$ denotes the pulse duration.
Then, the FDA signal transmitted by its $m$-th antenna can be expressed as
\begin{equation}\label{eq:sigmodel}
{{x}_{m}}\left( t \right)={{w}_{m}}s\left( t \right){{e}^{j2\pi {{f}_{m}}t}}
\end{equation}
where ${f_m} = {f_c} + \left( {m - 1} \right)\Delta f$ denotes the carrier frequency for the $m$-th antenna with $f_c$ and $\Delta f$ being the central frequency and FO, respectively.
Then, under the narrowband assumption, the synthesized signal seen at a specific location with a range-angle pair $\left( {r,\theta } \right)$ in the far field is given by 
\begin{equation}
\begin{aligned}
y\left( t \right)&=\sum\limits_{m=1}^{M}{{x_m}\left( t-\frac{{{r}_{m}}}{c} \right)} \\ 
& \approx s\left( t-\frac{r}{c} \right){{e}^{j2\pi {{f}_{c}}\left( t-\frac{r}{c} \right)}} \\ 
& \kern 9pt \times \mathbf{e}{{\left( t \right)}^{T}}\left( \mathbf{w}\odot {{\mathbf{a}}_{T,r}}\left( r \right)\odot {{\mathbf{a}}_{T,\theta }}\left( \Delta f,\theta  \right) \right) \\ 
\end{aligned}
\end{equation}
where ${r_m} = r - \left( {m - 1} \right)d\sin \theta $ denotes the slant range from the $m$-th transmit antenna to the target with $d$ being the inter-element spacing,
\begin{equation}
\mathbf{w}={{\left[ {{w}_{1}},{{w}_{2}},...,{{w}_{M}} \right]}^{T}}
\end{equation}
denotes the transmit beamforming vector, and
\begin{subequations}
	\begin{equation}
	\kern -6.6pt \mathbf{e}\left( t \right)={{\left[ 1,{{e}^{j2\pi \Delta ft}},...,{{e}^{j2\pi \left( M-1 \right)\Delta ft}} \right]}^{T}}
	\end{equation}
	\begin{equation}
	\kern -2pt {{\mathbf{a}}_{T,r}}\left( r \right)={{\left[ 1,{{e}^{-j2\pi \Delta f\frac{r}{c}}},...,{{e}^{-j2\pi \left( M-1 \right)\Delta f\frac{r}{c}}} \right]}^{T}}
	\end{equation}
	\begin{equation}
	\begin{aligned}
	& {{\mathbf{a}}_{T,{\theta }}}\left( \Delta f,\theta  \right)= \\ 
	& {{\left[ 1,{{e}^{j2\pi d\frac{\sin \theta }{c}\left( {{f}_{c}}+\Delta f \right)}},...,{{e}^{j2\pi d\frac{\sin \theta }{c}\left[ \left( M-1 \right){{f}_{c}}+{{\left( M-1 \right)}^{2}}\Delta f \right]}} \right]}^{T}}. \\ 
	\end{aligned}
	\end{equation}
\end{subequations}

Since the typical FDA transmit beampattern is time-variant \cite{wang2016overview}, by integrating it over the entire pulse duration, we define here the FDA integral transmit beampattern (FGTB) ${{P}_{\text{FGTB}}}\left( \mathbf{w},\Delta f,\theta  \right)$ for later use as 
\begin{equation}
\begin{aligned}
& {{P}_{\text{FGTB}}}\left( \mathbf{w},\Delta f,\theta  \right) \\ 
& =\int\limits_{r/c\ }^{r/c\ +{{T}_{P}}}{y\left( t \right){{y}^{c}}\left( t \right)\text{d}t} \\ 
& \kern -6pt \overset{{t}'=t-\frac{r}{c}}{\mathop{=}}\,\int\limits_{0}^{{{T}_{P}}}{\left\{ \begin{aligned}
	& \text{s}\left( {{t}'} \right){{\text{s}}^{c}}\left( {{t}'} \right){{\left( \mathbf{w}\odot {{\mathbf{a}}_{T,\theta }}\left( \Delta f,\theta  \right) \right)}^{H}}{{\mathbf{e}}^{c}}\left( {{t}'} \right) \\ 
	& \times {{\mathbf{e}}^{T}}\left( {{t}'} \right)\left( \mathbf{w}\odot {{\mathbf{a}}_{T,\theta }}\left( \Delta f,\theta  \right) \right) \\ 
	\end{aligned} \right\}\text{d}{t}'} \\ 
& =\operatorname{Tr}\left\{ {{\mathbf{R}}_{\text{FDA}}}\left( \mathbf{w}\odot {{\mathbf{a}}_{T,\theta }}\left( \Delta f,\theta  \right) \right){{\left( \mathbf{w}\odot {{\mathbf{a}}_{T,\theta }}\left( \Delta f,\theta  \right) \right)}^{H}} \right\} \\ 
\end{aligned}
\end{equation}
where 
\begin{equation}\label{eq:corr}
{{\mathbf{R}}_{\text{FDA}}}=\int\limits_{0}^{{{T}_{p}}}{s\left( {{t}'} \right){{\mathbf{e}}^{c}}\left( {{t}'} \right){{\mathbf{e}}^{T}}\left( {{t}'} \right){{s}^{c}}\left( {{t}'} \right)\operatorname{d}{t}'}
\end{equation}
can be seen as the correlation matrix of the FDA transmitted baseband signals.

According to the multi-channel FDA receiver proposed in \cite{gui2017coherent}, the echo signal $y(t)$ is first down-converted by multiple mixers with local carrier frequencies $\left\{ {{f}_{m}} \right\}_{m=1}^{M}$, respectively, and then matched filtered with $s(t)$. As a result, the $M \times N$-dimensional output ${{\mathbf{z}}_{\text{FDA}}}$ aligned to the target round trip delay $\frac{2r}{c}$ can be expressed as
\begin{equation}
\kern -2pt \begin{aligned}
& {{\mathbf{z}}_{\text{FDA}}} \\ 
& ={{\alpha }_{t}}\left[ \left( {{\mathbf{e}}^{c}}\left( t \right){{e}^{-j2\pi {{f}_{c}}t}}y\left( t-\frac{r}{c} \right) \right)*{{s}^{c}}\left( -t \right) \right]\left| _{t=\frac{2r}{c}} \right.\otimes {{\mathbf{a}}_{R,\theta }}\left( \theta  \right)  \\ 
& ={{\alpha }_{t}}\left[ {{\mathbf{R}}_{\text{FDA}}}\odot \left( {{\mathbf{a}}_{T,\theta }}\left( \Delta f,\theta  \right)\mathbf{a}_{T,r}^{T}\left( 2r \right) \right)\mathbf{w} \right]\otimes {{\mathbf{a}}_{R,\theta }}\left( \theta  \right)  \\ 
\end{aligned}
\end{equation}
where ${{\alpha }_{t}}$ is the target complex amplitude and
\begin{equation}
{{\mathbf{a}}_{R,\theta }}\left( \theta  \right)={{\left[ 1,{{e}^{j2\pi d\frac{\sin \theta }{c}{{f}_{c}}}},...,{{e}^{j2\pi \left( N-1 \right)d\frac{\sin \theta }{c}{{f}_{c}}}} \right]}^{T}}
\end{equation}
denotes the FDA receive steering vector.
Denote the $(p, q)$-th entry of ${{\mathbf{R}}_{\text{FDA}}}$ as ${{\mathbf{R}}_{\text{FDA}}}\left( p,q \right)$, such that
\begin{equation}
\begin{aligned}
{{\mathbf{R}}_{\text{FDA}}}\left( p,q \right)&=\int\limits_{0}^{{{T}_{p}}}{s\left( {{t}'} \right){{s}^{c}}\left( {{t}'} \right){{e}^{-j2\pi \left( p-1 \right)\Delta ft}}{{e}^{j2\pi \left( q-1 \right)\Delta ft}}\operatorname{d}{t}'} \\ 
& =\int\limits_{0}^{{{T}_{p}}}{s\left( {{t}'} \right){{s}^{c}}\left( {{t}'} \right){{e}^{j2\pi \left( q-p \right)\Delta ft}}\operatorname{d}{t}'}. \\ 
\end{aligned}
\end{equation}
Particularly, for the coherent FDA, the frequency offset $\Delta f$ is much smaller than the bandwidth of the baseband waveform $s(t)$, such that the correlation matrix in \eqref{eq:corr} can be approximated as \cite{gui2020adaptive} 
\begin{equation}
{{\mathbf{R}}_{\text{C-FDA}}}\approx {{\mathbf{1}}_{M\times M}},
\end{equation}
simplifying the output of coherent FDA as  
\begin{equation}
\begin{aligned}
{{\mathbf{z}}_{\text{C-FDA}}}&\approx {{\alpha }_{t}}\left[ {{\mathbf{1}}_{M\times M}}\odot \left( {{\mathbf{a}}_{T,\theta }}\left( \Delta f,\theta  \right)\mathbf{a}_{T,r}^{T}\left( 2r \right) \right)\mathbf{w} \right]\otimes {{\mathbf{a}}_{R,\theta }}\left( \theta  \right) \\ 
& ={{\alpha }_{t}}\left[ \left( {{\mathbf{a}}_{T,\theta }}\left( \Delta f,\theta  \right)\mathbf{a}_{T,r}^{T}\left( 2r \right) \right)\mathbf{w} \right]\otimes {{\mathbf{a}}_{R,\theta }}\left( \theta  \right) \\ 
& ={{\alpha }_{t}}\operatorname{vec}\left\{ {{\mathbf{a}}_{R,\theta }}\left( \theta  \right){{\mathbf{w}}^{T}}\left( {{\mathbf{a}}_{T,r}}\left( 2r \right)\mathbf{a}_{T,\theta }^{T}\left( \Delta f,\theta  \right) \right) \right\} \\ 
& ={{\alpha }_{t}}\mathbf{G}\left( \Delta f,r,\theta  \right)\mathbf{w} \\ 
\end{aligned}
\end{equation}
where
\begin{equation}
\mathbf{G}\left( \Delta f,r,\theta  \right)={{\mathbf{a}}_{R,\theta }}\left( \theta  \right)\otimes \left( {{\mathbf{a}}_{T,r}}\left( 2r \right)\mathbf{a}_{T,\theta }^{T}\left( \Delta f,\theta  \right) \right)
\end{equation}
and the FGTB of coherent FDA to
\begin{equation}\label{eq:fgtb}
{{P}_{\text{C-FGTB}}}\left( \mathbf{w},\Delta f,\theta  \right)\approx {{\mathbf{w}}^{H}}\mathbf{T}\left( \Delta f,\theta  \right)\mathbf{w}
\end{equation}
where
\begin{equation}\label{eq:ettta}
\mathbf{T}\left( \Delta f,\theta  \right)={{\mathbf{a}}_{T,\theta }}\left( \Delta f,\theta  \right)\mathbf{a}_{T,\theta }^{H}\left( \Delta f,\theta  \right).
\end{equation}

Assume that the output of coherent FDA is filtered through the $MN\times 1$-dimensional receive weight vector $\bf{b}$, and then the received RPDE $\eta \left( \mathbf{w},\mathbf{b} \right)$ can be expressed as
\begin{equation}\label{eq:eta}
\begin{aligned}
\eta \left( \mathbf{w},\mathbf{b} \right)&=\frac{\oint_{{{\Delta }_{d}}\left( r,\theta  \right)}{{{\left| {{\mathbf{b}}^{H}}{{\mathbf{z}}_{\text{C-FDA}}} \right|}^{2}}\operatorname{d}r\operatorname{d}\theta }}{\oint_{\Delta \left( r,\theta  \right)}{{{\left| {{\mathbf{b}}^{H}}{{\mathbf{z}}_{\text{C-FDA}}} \right|}^{2}}\operatorname{d}r\operatorname{d}\theta }} \\ 
& =\frac{\oint_{{{\Delta }_{d}}\left( r,\theta  \right)}{{{\left| {{\mathbf{b}}^{H}}\mathbf{G}\left( \Delta f,r,\theta  \right)\mathbf{w} \right|}^{2}}\operatorname{d}r\operatorname{d}\theta }}{\oint_{\Delta \left( r,\theta  \right)}{{{\left| {{\mathbf{b}}^{H}}\mathbf{G}\left( \Delta f,r,\theta  \right)\mathbf{w} \right|}^{2}}\operatorname{d}r\operatorname{d}\theta }} \\ 
\end{aligned}
\end{equation}
where 
\begin{equation}
{{\Delta }_{d}}\left( r,\theta  \right):\left\{ \begin{matrix}
\theta \in {{\Theta }_{d}}  \\
r\in {{R}_{d}}  \\
\end{matrix} \right.
\end{equation}
and
\begin{equation}
\Delta \left( r,\theta  \right):\left\{ \begin{matrix}
-\frac{\pi }{2}\le \theta \le \frac{\pi }{2}  \\
{{r}_{min }}\le r\le {{r}_{max }}  \\
\end{matrix} \right.
\end{equation}
denote the two-dimensional range-angle region of interest and the whole potential region, respectively.
$r_{min}$ and $r_{max}$ denote the minimum and maximum radar distances of interest, respectively, which are determined by the radar range gate in practice.
${{\Theta }_{d}}$ and ${{R}_{d}}$ denote the interested direction and distance regions, respectively.
 
\section{Problem formulation}\label{sec:b}
\noindent 
The goal of this paper is to jointly optimize the transmit weight vector ${\mathbf{w}}$ and the receive weight vector ${\mathbf{b}}$ such that the received RPDE is maximized.

\subsection{Similarity constraint}
\noindent
Similarity constraint has been widely used in MIMO radar waveform design \cite{cui2013mimo,cheng2017mimo,aldayel2016successive}.
For coherent FDA considered in this paper, enforcing a similarity constraint on the transmit weight vector allows a compromise between optimizing the received RPDE and controlling the transmit power allocation.
Specifically, for the transmission side of the coherent FDA, we are interested in focusing the transmit energy on the potential spatial section of interest for coherent gain. 
To this end, the transmit weight vector ${\mathbf{w}}$ should be designed to maximize the mainlobe-to-sidelobe ratio of the FGTB of coherent FDA derived in \eqref{eq:fgtb}, formulated as
\begin{equation}
{{P}_{1}}\left\{ \begin{matrix}
\underset{\mathbf{w}}{\mathop{\max }}\, & \frac{\int_{{{\Theta }_{d}}}{{{\mathbf{w}}^{H}}\mathbf{T}\left( \Delta f,\theta  \right)\mathbf{w}\operatorname{d}\theta }}{\int_{-\frac{\pi }{2}}^{\frac{\pi }{2}}{{{\mathbf{w}}^{H}}\mathbf{T}\left( \Delta f,\theta  \right)\mathbf{w}\operatorname{d}\theta }}=\frac{{{\mathbf{w}}^{H}}{{\mathbf{\Omega }}_{d}}\left( \Delta f \right)\mathbf{w}}{{{\mathbf{w}}^{H}}\mathbf{\Omega }\left( \Delta f \right)\mathbf{w}}  \\
\end{matrix} \right.
\end{equation}
where
\begin{subequations}
	\begin{equation}
{{\mathbf{\Omega }}_{d}}\left( \Delta f \right)=\int_{{{\Theta }_{d}}}{\mathbf{T}\left( \Delta f,\theta  \right)\operatorname{d}\theta }
	\end{equation}
	\begin{equation}
	\mathbf{\Omega }\left( \Delta f \right)=\int_{-\frac{\pi }{2}}^{\frac{\pi }{2}}{\mathbf{T}\left( \Delta f,\theta \right)\operatorname{d}\theta }.
	\end{equation}
\end{subequations}
It is worth noting that since the derived FGTB is range-independent, the emitted energy can only be concentrated in the angle domain by the design.  However, the output ${{\mathbf{z}}_{\text{C-FDA}}}$ of the multi-channel coherent FDA receiver is a angle- and range-dependent vector, the received energy can be focused in the angle-range plane by jointly designing the transmit and receive weights.
The optimal solution ${\mathbf{ w}}_0$ to the problem $P_1$ is given by
\begin{equation}
{{\mathbf{w}}_{0}}={{\lambda }_{\max ,vec}}\left\{ {{\mathbf{\Omega }}^{-1}}\left( \Delta f \right){{\mathbf{\Omega }}_{d}}\left( \Delta f \right) \right\},
\end{equation}
as shown in the \ref{sec:appen}.
As a result, the similarity constraint has the form
\begin{equation}\left\| {{\mathbf{w}} - \rho {{\mathbf{w}}_0}} \right\| \leqslant 1,\kern 20pt {\rho^2} \leqslant 1
\end{equation}
where $\rho$ is a scaling parameter such that the reference weight ${\mathbf{w}}_0$ is chosen according to the
actual norm of the designed transmit weight.
This is equivalent to optimizing the received RPDE in a suitable neighborhood of a reference transmit weight vector ${\mathbf{ w}}_0$, which is known to have good properties.
\subsection{Energy constraint}
\noindent
Since the FDA essentially transmits a multi-carrier signal, it is expected that the transmit energy of each antenna should be no larger than a threshold \cite{de2011design,richards2014fundamentals}. 
From \eqref{eq:sigmodel}, the energy radiated by the $m$-th antenna can be expressed as
\begin{equation}
\int_{{{T}_{p}}}{{{\left| {{w}_{m}}s\left( t \right){{e}^{j2\pi {{f}_{m}}t}} \right|}^{2}}\operatorname{d}t}={{\left| {{w}_{m}} \right|}^{2}}\int_{{{T}_{p}}}{{{\left| s\left( t \right) \right|}^{2}}\operatorname{d}t}={{\left| {{w}_{m}} \right|}^{2}}.
\end{equation}
Without loss of generality, assuming that the maximum energy of each transmit antenna to $\frac{1}{M}$, the energy constraint can thus be written as
\begin{equation}
{{\mathbf{w}}^H}{{\mathbf{E}}_m}{\mathbf{w}} \leqslant \frac{1}{M} 
\end{equation}
for $m = 1,...,M - 1$, where 
\begin{equation}
{{\mathbf{E}}_{m}}\left( i,j \right)=\left\{ \begin{matrix}
1 & i=m,j=m  \\
0 & otherwise  \\
\end{matrix} \right..
\end{equation}

\subsection{Optimization problem}
\noindent
Based on the aforementioned discussions, aiming to
maximize the received RPDE, the joint design of transmit and receive
weight vectors for coherent FDA can be formulated as
\begin{equation}
{{P}_{2}}\left\{ \begin{matrix}
\underset{\mathbf{w},\mathbf{b}}{\mathop{\max }}\, & \eta \left( \mathbf{w},\mathbf{b} \right)  \\
\operatorname{s}.t. & \left\{ \begin{matrix}
\left\| \mathbf{w}-\rho {{\mathbf{w}}_{0}} \right\|\le 1,{{\rho }^{2}}\le 1  \\
{{\mathbf{w}}^{H}}{{\mathbf{E}}_{m}}\mathbf{w}\le \frac{1}{M},m=1,...,M-1  \\
\end{matrix} \right.  \\
\end{matrix} \right.
\end{equation}
where $\eta \left( \mathbf{w},\mathbf{b} \right) $ is given in \eqref{eq:eta}.
In order to obtain the integration value of the objective function $\eta \left( \mathbf{w},\mathbf{b} \right)$, we adopt a sufficient approximation technique by choosing dense grid points in the whole observation region. The approximation is written as
\begin{equation}
\eta \left( \mathbf{w},\mathbf{b} \right)\approx \frac{\sum\limits_{{{t}_{1}=1}}^{{{T}_{1}}}{\sum\limits_{{{t}_{2}=1}}^{{{T}_{2}}}{{{\left| {{\mathbf{b}}^{H}}\mathbf{G}\left( \Delta f,{{r}_{{{t}_{1}}}},{{\theta }_{{{t}_{2}}}} \right)\mathbf{w} \right|}^{2}}}}}{\sum\limits_{g=1}^{{{J}_{1}}}{\sum\limits_{h=1}^{{{J}_{2}}}{{{\left| {{\mathbf{b}}^{H}}\mathbf{G}\left( \Delta f,{{r}_{g}},{{\theta }_{h}} \right)\mathbf{w} \right|}^{2}}}}}
\end{equation}
where $T_1$ and $T_2$ denote the number of grid points in the range and angle dimensions of the desired area ${{\Delta }_{d}}\left( r,\theta  \right)$, respectively. $J_1$ and $J_2$ denote the number of grid points in the range and angle dimensions of the whole observation region $\Delta \left( r,\theta  \right)$, respectively. 
Thus, following the results in \cite{aubrry2016forcing}, the problem $P_2$ can be converted to
\begin{equation}
{{P}_{3}}\left\{ \begin{matrix}
\underset{\mathbf{w},\mathbf{b}}{\mathop{\max }}\, & \frac{\sum\limits_{{{t}_{1}}=1}^{{{T}_{1}}}{\sum\limits_{{{t}_{2}}=1}^{{{T}_{2}}}{{{\left| {{\mathbf{b}}^{H}}\mathbf{G}\left( \Delta f,{{r}_{{{t}_{1}}}},{{\theta }_{{{t}_{2}}}} \right)\mathbf{w} \right|}^{2}}}}}{\sum\limits_{g=1}^{{{J}_{1}}}{\sum\limits_{h=1}^{{{J}_{2}}}{{{\left| {{\mathbf{b}}^{H}}\mathbf{G}\left( \Delta f,{{r}_{g}},{{\theta }_{h}} \right)\mathbf{w} \right|}^{2}}}}}  \\
\operatorname{s}.t. & \left\{ \begin{matrix}
{{\mathbf{w}}^{H}}\mathbf{\Gamma }\left( {{\mathbf{w}}_{0}} \right)\mathbf{w}\le 1  \\
{{\mathbf{w}}^{H}}{{\mathbf{E}}_{m}}\mathbf{w}\le \frac{1}{M},m=1,...,M  \\
\end{matrix} \right.  \\
\end{matrix} \right.
\end{equation}
where ${\mathbf{\Gamma }}\left( {{{\mathbf{w}}_0}} \right) = {{\mathbf{I}}_M} - {{\mathbf{w}}_0}{\mathbf{w}}_0^H$.
Since the objective function is a quadratic fractional function with respect to ${\mathbf{w}}$ and ${\mathbf{b}}$, problem $P_3$ is nonconvex and NP-hard problem \cite{boyd2004convex}.
In the following, a sequential optimization algorithm is devised, capable of yielding the optimal parameter pair with good performance.

\section{Optimization algorithm implementation}\label{sec:c}
\noindent
In order to relax problem $P_3$, we separate the maximization,
such that this is done alternatively with respect to $\bf{w}$
and $\bf{b}$, keeping the other fixed. 
Note that sequential optimization algorithms are guaranteed to converge and achieve the best balance between performance and complexity \cite{8454321,9693236}.
\subsection{Optimization of ${\mathbf{b}}$}
\noindent
For a fixed transmit weight vector ${\mathbf{w}}$, the optimal receive weight vector can be obtained by rewriting $P_3$ as
\begin{equation}
\begin{matrix}
\underset{\mathbf{b}}{\mathop{\max }}\, & \frac{\sum\limits_{{{t}_{1}}=1}^{{{T}_{1}}}{\sum\limits_{{{t}_{2}}=1}^{{{T}_{2}}}{{{\left| {{\mathbf{b}}^{H}}\mathbf{G}\left( \Delta f,{{r}_{{{t}_{1}}}},{{\theta }_{{{t}_{2}}}} \right)\mathbf{w} \right|}^{2}}}}}{\sum\limits_{g=1}^{{{J}_{1}}}{\sum\limits_{h=1}^{{{J}_{2}}}{{{\left| {{\mathbf{b}}^{H}}\mathbf{G}\left( \Delta f,{{r}_{g}},{{\theta }_{h}} \right)\mathbf{w} \right|}^{2}}}}}  \\
\end{matrix}.
\end{equation}
It can be simplified to a similar form as problem $P_1$, given by
\begin{equation}
{{P}_{4}}\left\{ \begin{matrix}
\underset{\mathbf{b}}{\mathop{\max }}\, & \frac{{{\mathbf{b}}^{H}}{\bf{\Pi }_{d}}\left( \mathbf{w} \right)\mathbf{b}}{{{\mathbf{b}}^{H}}\bf{\Pi} \left( \mathbf{w} \right)\mathbf{b}}  \\
\end{matrix} \right.
\end{equation}
where
\begin{subequations}
	\begin{equation}
	\boldsymbol{{\Pi }_{d}}\left( \mathbf{w} \right)=\sum\limits_{{{t}_{1}}=1}^{{{T}_{1}}}{\sum\limits_{{{t}_{2}}=1}^{{{T}_{2}}}{\mathbf{G}\left( \Delta f,{{r}_{{{t}_{1}}}},{{\theta }_{{{t}_{2}}}} \right)\mathbf{w}{{\mathbf{w}}^{H}}{{\mathbf{G}}^{H}}\left( \Delta f,{{r}_{{{t}_{1}}}},{{\theta }_{{{t}_{2}}}} \right)}}
	\end{equation}
	\begin{equation}
	\kern -3.5pt \boldsymbol{\Pi} \left( \mathbf{w} \right)=\sum\limits_{g=1}^{{{J}_{1}}}{\sum\limits_{h=1}^{{{J}_{2}}}{\mathbf{G}\left( \Delta f,{{r}_{g}},{{\theta }_{h}} \right)\mathbf{w}{{\mathbf{w}}^{H}}{{\mathbf{G}}^{H}}\left( \Delta f,{{r}_{g}},{{\theta }_{h}} \right)}}.
	\end{equation}
\end{subequations}
The analytical solution of problem $P_4$ can then be obtained as
\begin{equation}
{{\mathbf{b}}^{*}}={{\lambda }_{\max ,vec}}\left\{ {\boldsymbol{\Pi }^{-1}}\left( \mathbf{w} \right){\boldsymbol{\Pi }_{d}}\left( \mathbf{w} \right) \right\}.
\end{equation}

\subsection{Optimization of ${\mathbf{w}}$}
\noindent
After obtaining the receive weight vector $\bf{b}$, we now solve for $\bf{w}$ with fixed $\bf{b}$.
Maximizing the ratio is equivalent to maximizing the numerator while fixing the denominator, recasting the probelm $P_3$ as
\begin{equation}
{{P}_{4}}\left\{ \begin{matrix}
\underset{\mathbf{w}}{\mathop{\max }}\, & \sum\limits_{g=1}^{{{J}_{1}}}{\sum\limits_{h=1}^{{{J}_{2}}}{{{\left| {{\mathbf{b}}^{H}}\mathbf{G}\left( \Delta f,{{r}_{g}},{{\theta }_{h}} \right)\mathbf{w} \right|}^{2}}}}  \\
\operatorname{s}.t. & \left\{ \begin{matrix}
{{\mathbf{w}}^{H}}\mathbf{\Gamma }\left( {{\mathbf{w}}_{0}} \right)\mathbf{w}\le 1  \\
{{\mathbf{w}}^{H}}{{\mathbf{E}}_{m}}\mathbf{w}\le \frac{1}{M},m=1,...,M  \\
\sum\limits_{{{t}_{1}}=1}^{{{T}_{1}}}{\sum\limits_{{{t}_{2}}=1}^{{{T}_{2}}}{{{\left| {{\mathbf{b}}^{H}}\mathbf{G}\left( \Delta f,{{r}_{{{t}_{1}}}},{{\theta }_{{{t}_{2}}}} \right)\mathbf{w} \right|}^{2}}}}=1  \\
\end{matrix} \right.  \\
\end{matrix} \right.
\end{equation}
Before proceeding, let us rewrite the problem $P_4$ as
\begin{equation}
{{P}_{5}}\left\{ \begin{matrix}
\underset{\mathbf{w}}{\mathop{\max }}\, & {{\mathbf{w}}^{H}}\boldsymbol{\Xi }_{d}\left( \mathbf{b} \right)\mathbf{w}  \\
\operatorname{s}.t. & \left\{ \begin{matrix}
{{\mathbf{w}}^{H}}\mathbf{\Gamma }\left( {{\mathbf{w}}_{0}} \right)\mathbf{w}\le 1  \\
{{\mathbf{w}}^{H}}{{\mathbf{E}}_{m}}\mathbf{w}\le \frac{1}{M},m=1,...,M  \\
{{\mathbf{w}}^{H}}{{\mathbf{\Xi }}}\left( \mathbf{b} \right)\mathbf{w}=1  \\
\end{matrix} \right.  \\
\end{matrix} \right.
\end{equation}
where
\begin{subequations}
	\begin{equation}
	\kern -3.5pt {{\mathbf{\Xi }}_{d}}\left( \mathbf{b} \right)=\sum\limits_{{{t}_{1}}=1}^{{{T}_{1}}}{\sum\limits_{{{t}_{2}}=1}^{{{T}_{2}}}{\boldsymbol{\gamma }\left( \mathbf{b};{{r}_{{{t}_{1}}}},{{\theta }_{{{t}_{2}}}} \right){{\boldsymbol{\gamma }}^{H}}\left( \mathbf{b};{{r}_{{{t}_{1}}}},{{\theta }_{{{t}_{2}}}} \right)}}
	\end{equation}
	\begin{equation}
	\mathbf{\Xi }\left( \mathbf{b} \right)=\sum\limits_{g=1}^{{{J}_{1}}}{\sum\limits_{h=1}^{{{J}_{2}}}{\boldsymbol{\gamma }\left( \mathbf{b};{{r}_{{{t}_{1}}}},{{\theta }_{{{t}_{2}}}} \right){{\boldsymbol{\gamma }}^{H}}\left( \mathbf{b};{{r}_{{{t}_{1}}}},{{\theta }_{{{t}_{2}}}} \right)}}
	\end{equation}
\end{subequations}
where
\begin{equation}
\boldsymbol{\gamma }\left( \mathbf{b};r,\theta  \right)={{\mathbf{G}}^{H}}\left( \Delta f,r,\theta  \right)\mathbf{b}.
\end{equation}
Obviously, the problem $P_5$ is a nonconvex QCQP (quadratically constrained quadratic program) problem and no closed-form solution exists.
Introducing a new variable ${\mathbf{W}}$ and noting that  ${\mathbf{W}} = {\mathbf{w}}{{\mathbf{w}}^H}$ is equivalent to ${\mathbf{W}}$ being a symmetric positive semidefinite (PSD) matrix of rank one, we have
\begin{equation}
{{P}_{6}}\left\{ \begin{matrix}
\underset{\mathbf{W}}{\mathop{\min }}\, & \operatorname{Tr}\left\{ \mathbf{\Xi }_{d}\left( \mathbf{b} \right)\mathbf{W} \right\}  \\
\operatorname{s}.t. & \begin{matrix}
\operatorname{Tr}\left\{ \mathbf{\Gamma }\left( {{\mathbf{w}}_{0}} \right)\mathbf{W} \right\}\le 1  \\
\operatorname{Tr}\left\{ {{\mathbf{E}}_{m}}\mathbf{W} \right\}\frac{1}{M},m=1,...,M-1  \\
\operatorname{Tr}\left\{ {{\mathbf{\Xi }}}\left( \mathbf{b} \right)\mathbf{W} \right\}=1  \\
\mathbf{W}\succeq 0  \\
\operatorname{Rank}\left\{ \mathbf{W} \right\}=1  \\
\end{matrix}  \\
\end{matrix} \right.
\end{equation}
It can be seen that problem $P_6$ can be easily solved by applying  the SDR. In particular, SDR is a powerful, computationally efficient technique and is capable of providing accurate approximation for the nonconvex QCQPs in an almost mechanical fashion \cite{luo2010semidefinite}. 

First, we drop the nonconvex rank constraint to obtain the following relaxed version of $P_6$:
\begin{equation}
{{P}_{7}}\left\{ \begin{matrix}
\underset{\mathbf{W}}{\mathop{\min }}\, & \operatorname{Tr}\left\{ \mathbf{\Xi }_{d}\left( \mathbf{b} \right)\mathbf{W} \right\}  \\
\operatorname{s}.t. & \begin{matrix}
\operatorname{Tr}\left\{ \mathbf{\Gamma }\left( {{\mathbf{w}}_{0}} \right)\mathbf{W} \right\}\le 1  \\
\operatorname{Tr}\left\{ {{\mathbf{E}}_{m}}\mathbf{W} \right\}\frac{1}{M},m=1,...,M-1  \\
\operatorname{Tr}\left\{ {{\mathbf{\Xi }}}\left( \mathbf{b} \right)\mathbf{W} \right\}=1  \\
\mathbf{W}\succeq 0  \\
\end{matrix}  \\
\end{matrix} \right.
\end{equation}
Problem $P_7$ is a convex semidefinite programming (SDP).
A solution may be obtained in polynomial time using standard optimization tools, such as the convex optimization toolbox CVX  \cite{grant2014cvx}.
If the rank of the globally optimal solution ${{\mathbf{W}}^{*}}$ obtained for problem $P_7$ is one, then ${{\mathbf{W}}^{*}}={{\mathbf{w}}^{*}}{{\left( {{\mathbf{w}}^{*}} \right)}^{H}}$, with ${{\mathbf{w}}^{*}}$ being the globally optimal solution to problem $P_4$. 
However, in most cases, the SDP solution is infeasible for problem $P_6$.
Then, we need to extract an optimal nonconvex QCQP solution from the SDR solution.
Particularly, the randomization method has been empirically found to yield promising approximations if sufficient number of randomization trials is employed \cite{luso2010sdp}.
The proposed iterative algorithm for problem $P_3$ is summarized in Algorithm \ref{alg:1}.

\begin{breakablealgorithm} 
	\renewcommand{\algorithmicrequire}{\textbf{Input:}}
	\renewcommand{\algorithmicensure}{\textbf{Output:}}
	\caption{The Proposed Iterative Algorithm}
    \label{alg:1} 
	\begin{algorithmic}[1] 
		\REQUIRE Spatial sections of interest ${{\Delta }_{d}}\left( r,\theta  \right)$, the whole potential region ${{\Delta }}\left( r,\theta  \right)$, and the number of iterations ${{G}_{max }}$.\\ 
		\ENSURE The optimal transmit and receive weight vectors $\left( {{{\mathbf{w}}^ * },{{\mathbf{b}}^ * }} \right)$.
		\STATE Compute \\
		$\kern 45pt {{\mathbf{\Omega }}_{d}}\left( \Delta f \right)=\int_{{{\theta }_{1}}}^{{{\theta }_{2}}}{\mathbf{T}\left( \Delta f,\theta  \right)\operatorname{d}\theta }$,\\
		$\kern 50pt \mathbf{\Omega }\left( \Delta f \right)=\int_{-\frac{\pi }{2}}^{\frac{\pi }{2}}{\mathbf{T}\left( \Delta f,\theta  \right)\operatorname{d}\theta }$.\\
		Obtain the reference transmit weight vector\\ $\kern 35pt {{\mathbf{w}}_{0}}={{\lambda }_{max,vec}}\left\{ {{\mathbf{\Omega }}^{-1}}\left( \Delta f \right){{\mathbf{\Omega }}_{d}}\left( \Delta f \right) \right\}$.
		\STATE For $q = 1$, initialize the transmit weight vector ${{\mathbf{w}}_1} = {{\mathbf{w}}_0}$.  
		\STATE Let $q = q + 1$, compute \\
		$\kern 6pt {{\mathbf{\Pi }}_{\mathbf{d}}}\left( {{\mathbf{w}}_{q-1}} \right)=\sum\limits_{{{t}_{1}}=1}^{{{T}_{1}}}{\sum\limits_{{{t}_{2}}=1}^{{{T}_{2}}}{\left\{ \begin{aligned}
				& \mathbf{G}\left( \Delta f,{{r}_{{{t}_{1}}}},{{\theta }_{{{t}_{2}}}} \right){{\mathbf{w}}_{q-1}} \\ 
				& \times \mathbf{w}_{q-1}^{H}{{\mathbf{G}}^{H}}\left( \Delta f,{{r}_{{{t}_{1}}}},{{\theta }_{{{t}_{2}}}} \right) \\ 
				\end{aligned} \right\}}}$,\\
		$\kern 12pt \mathbf{\Pi }\left( {{\mathbf{w}}_{q-1}} \right)=\sum\limits_{g=1}^{{{J}_{1}}}{\sum\limits_{h=1}^{{{J}_{2}}}{\left\{ \begin{aligned}
				& \mathbf{G}\left( \Delta f,{{r}_{g}},{{\theta }_{h}} \right){{\mathbf{w}}_{q-1}} \\ 
				& *\mathbf{w}_{q-1}^{H}{{\mathbf{G}}^{H}}\left( \Delta f,{{r}_{g}},{{\theta }_{h}} \right) \\ 
				\end{aligned} \right\}}}$.\\
		Obtain \\
		$\kern 30pt {{\mathbf{b}}_{q-1}}={{\lambda }_{\max ,vec}}\left\{ \mathbf{\Pi }{{\left( {{\mathbf{w}}_{q-1}} \right)}^{-1}}{{\mathbf{\Pi }}_{\mathbf{d}}}\left( {{\mathbf{w}}_{q-1}} \right) \right\}$.
		\STATE Compute\\
		$\kern -5pt {{\mathbf{\Xi }_{d}}}\left( {{\mathbf{b}}_{q-1}} \right)=\sum\limits_{{{t}_{1}}=1}^{{{T}_{1}}}{\sum\limits_{{{t}_{2}}=1}^{{{T}_{2}}}{\boldsymbol{\gamma }\left( {{\mathbf{b}}_{q-1}};{{r}_{{{t}_{1}}}},{{\theta }_{{{t}_{2}}}} \right){{\boldsymbol{\gamma }}^{H}}\left( {{\mathbf{b}}_{q-1}};{{r}_{{{t}_{1}}}},{{\theta }_{{{t}_{2}}}} \right)}}$,\\
		$\mathbf{\Xi }\left( {{\mathbf{b}}_{q-1}} \right)=\sum\limits_{g=1}^{{{J}_{1}}}{\sum\limits_{h=1}^{{{J}_{2}}}{\boldsymbol{\gamma }\left( {{\mathbf{b}}_{q-1}};{{r}_{{{t}_{1}}}},{{\theta }_{{{t}_{2}}}} \right){{\boldsymbol{\gamma }}^{H}}\left( {{\mathbf{b}}_{q-1}};{{r}_{{{t}_{1}}}},{{\theta }_{{{t}_{2}}}} \right)}}$.\\
		Solve the SDP problem below and denote ${\mathbf{\hat W}}$ as the solution of\\
		$\kern 10pt \begin{matrix}
		\underset{\mathbf{W}}{\mathop{\min }}\, & \operatorname{Tr}\left\{ \mathbf{\Xi }_{d}\left( {{\mathbf{b}}_{q-1}} \right)\mathbf{W} \right\}  \\
		\operatorname{s}.t. & \begin{matrix}
		\operatorname{Tr}\left\{ \mathbf{\Gamma }\left( {{\mathbf{w}}_{0}} \right)\mathbf{W} \right\}\le 1  \\
		\operatorname{Tr}\left\{ {{\mathbf{E}}_{m}}\mathbf{W} \right\}\le \frac{1}{M},m=1,...,M-1  \\
		\operatorname{Tr}\left\{ {{\mathbf{\Xi }}}\left( {{\mathbf{b}}_{q-1}} \right)\mathbf{W} \right\}=1  \\
		\mathbf{W}\succeq \mathbf{0}  \\
		\end{matrix}  \\
		\end{matrix}$
		\STATE Perform the eigenvalue decomposition of ${\mathbf{\hat W}}$, i.e. ${\mathbf{\hat W}} = {\mathbf{V\Delta }}{{\mathbf{V}}^H}$.
		\STATE If $\operatorname{Rank} \left\{ {\mathbf{V}} \right\} = 1$, then ${\mathbf{V}} = {\mathbf{v}}{{\mathbf{v}}^H}$, output ${{\mathbf{w}}_q} = {\mathbf{v}}$  and skip to Step 8.
		\STATE  Run randomization method steps as follows:\\
		$\bullet$ Perform the Cholesky factorization ${\mathbf{\hat W}} = {\mathbf{R}}_{\mathbf{w}}^H{{\mathbf{R}}_{\mathbf{w}}}$;\\
		$\bullet$ Generate $K$ independent identically distributed normal Gaussian random vectors ${\boldsymbol{\sigma}_k},k = 1,2,...,K$ with $K$ being the number of the randomization trials;\\
		$\bullet$ For the $k$-th randomization trial, construct new random vector ${{\boldsymbol{\xi }}_k} = {\mathbf{R}}_{\mathbf{w}}^H{\boldsymbol{\sigma} _k}$;\\
		$\bullet$ Apply a rescaling method \\
		$\kern 32pt {{\boldsymbol{\varsigma }}_k} = \frac{{{{\boldsymbol{\xi }}_k}}}{{\sqrt {{\boldsymbol{\xi }}_k^H{\boldsymbol{\gamma }}\left( {{{\mathbf{b}}_{q - 1}};\theta ,r} \right){{\boldsymbol{\gamma }}^H}\left( {{{\mathbf{b}}_{q - 1}};\theta ,r} \right){{\boldsymbol{\xi }}_k}} }}$;\\
		$\bullet$ Select ${\boldsymbol{\vartheta}_l}, l = 1,2,...,L$ from set ${\mathbb{R}_1}\left( {{{\boldsymbol{\varsigma }}_k}} \right) \cap {\mathbb{R}_2}\left( {{{\boldsymbol{\varsigma }}_k}} \right)$; \\
		$\kern 35pt {\mathbb{R}_1}\left( {{{\boldsymbol{\varsigma }}_k}} \right) = \left\{ {{{\boldsymbol{\varsigma }}_k}\left| {{\boldsymbol{\varsigma }}_k^H{\boldsymbol{\Gamma }}\left( {{{\mathbf{w}}_0}} \right){{\boldsymbol{\varsigma }}_k} \leqslant 1} \right.} \right\}$\\
		$\kern 35pt {\mathbb{R}_2}\left( {{{\boldsymbol{\varsigma }}_k}} \right) = \left\{ {{{\boldsymbol{\varsigma }}_k}\left| {{\boldsymbol{\varsigma }}_k^H{{\mathbf{E}}_m}{{\boldsymbol{\varsigma }}_k} \leqslant \frac{1}{M}} \right.} \right\}$ \\
		for $ m = 1,...,M - 1$ with $L$ being the number of the feasible trials;\\ 
		$\bullet$ Choose ${{\mathbf{w}}_q} = {\boldsymbol{\vartheta} ^ * }$ such that\\
		$\kern 40pt{\boldsymbol{\vartheta} ^ * } = \arg \mathop {{\text{min}}}\limits_{{\boldsymbol{\vartheta} _l}} \boldsymbol{\vartheta} _l^H\boldsymbol{\Xi} \left( {{{\mathbf{b}}_{q - 1}}} \right){\boldsymbol{\vartheta} _l}$;
		\STATE Skip to Step 3 until $q=G_{max}$. 
		\STATE Output ${{\mathbf{w}}^ * } = {{\mathbf{w}}_{G_{max}}}$ and ${{\mathbf{b}}^ * } = {{\mathbf{b}}_{G_{max}-1}}$.
	\end{algorithmic}  
\end{breakablealgorithm}

\section{Simulation}\label{sec:d}
\noindent
In this section, we provide various numerical simulations to examine the performance of the designed transmit-receive scheme for coherent FDA systems. 
Specifically, in the simulations, we consider two desired scenarios, i.e., single-target and multi-target.
The assumed system parameters are listed in Table \ref{table:tabb} and the entire area is $\Delta \left( r,\theta  \right):\left\{ {\begin{array}{*{20}{c}}
	{ - {{90}^o} \leqslant \theta  \leqslant {{90}^o}} \\ 
	{60\kern 1pt km \leqslant r \leqslant 85\kern 1ptkm} 
	\end{array}} \right.$. 
For the proposed algorithm, the number of iterations and randomization trails are $G_{max}=20$ and $K = 200$, respectively.

\begin{table}
	\centering
	\caption{System parameters}
	{\begin{tabular}[l]{@{}ll}
			\toprule
			Parameter & Value\\
			\midrule
			Number of transmit antennas &$\kern -0.4pt M = 10$ \\
			Number of receive antennas &$\kern 0.8pt N = 10$ \\
			Inter-element spacing & $\kern 4.5pt d = 1.5 \kern 2pt cm $\\ 
			Frequency offset & $\kern -4.5pt \Delta f=5 \kern 2pt kHz$\\
			Pulse duration & $T_p = 10 \kern 2pt\mu s$\\
			Carrier frequency&$\kern 1.6pt f_c = 10 \kern 2pt GHz$ \\
			Bandwidth of baseband waveform & $\kern 5pt 70 \kern 2pt MHz$\\
			\bottomrule
	\end{tabular}}
\label{table:tabb}
\end{table}

\subsection{Single target}
\noindent
We assume the potential spatial sections of interest for the transmitter are ${{\Theta }_{d}}=\left[ {{40}^{o}},{{60}^{o}} \right]$, and the target is located at $\left( {{{50}^o},80\kern 1pt km} \right)$ with a mesh grid size of $\left( {{{0.2}^o},0.1\kern 1pt km} \right)$ for integral approximation.

\begin{figure}[t]
	\centering
	\includegraphics[width=0.5\textwidth]{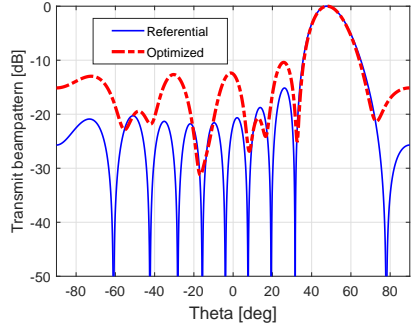}
	\caption{Comparison between the optimized transmit beampattern and the reference transmit beampattern (single target).}\label{fig:aa}
\end{figure}

Figure~\ref{fig:aa} compares the transmit beampattern obtained with the reference transmit weight vector ${{\mathbf{w}}_{0}}$ and the transmit beampattern designed with the optimized transmit weight vector ${{\mathbf{w}}^{*}}$.
The beampattern $P\left( \theta  \right)$ is computed as
\begin{equation}
P\left( \theta  \right)={{\mathbf{w}}^{H}}\mathbf{T}\left( \Delta f,\theta  \right)\mathbf{w}
\end{equation}
where $\mathbf{T}\left( \Delta f,\theta  \right)$ is given in \eqref{eq:ettta}.
The result reveals that the designed transmit beampattern has wider mainlobe and higher sidelobe.
In order to achieve energy focusing at the coherent FDA receiver, the transmit and receive weights need to cooperate with each other, which may compromise the performance of transmit beam pattern for optimal results.

\begin{figure}[t]
	\centering
	\includegraphics[width=0.5\textwidth]{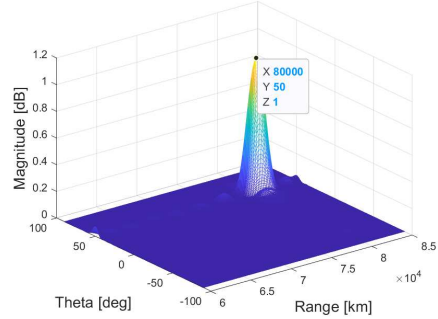}
	\caption{Energy distribution of the coherent FDA radar in receiver end.}\label{fig:bb}
\end{figure}

\begin{figure}[t]
	\centering
	\includegraphics[width=0.5\textwidth]{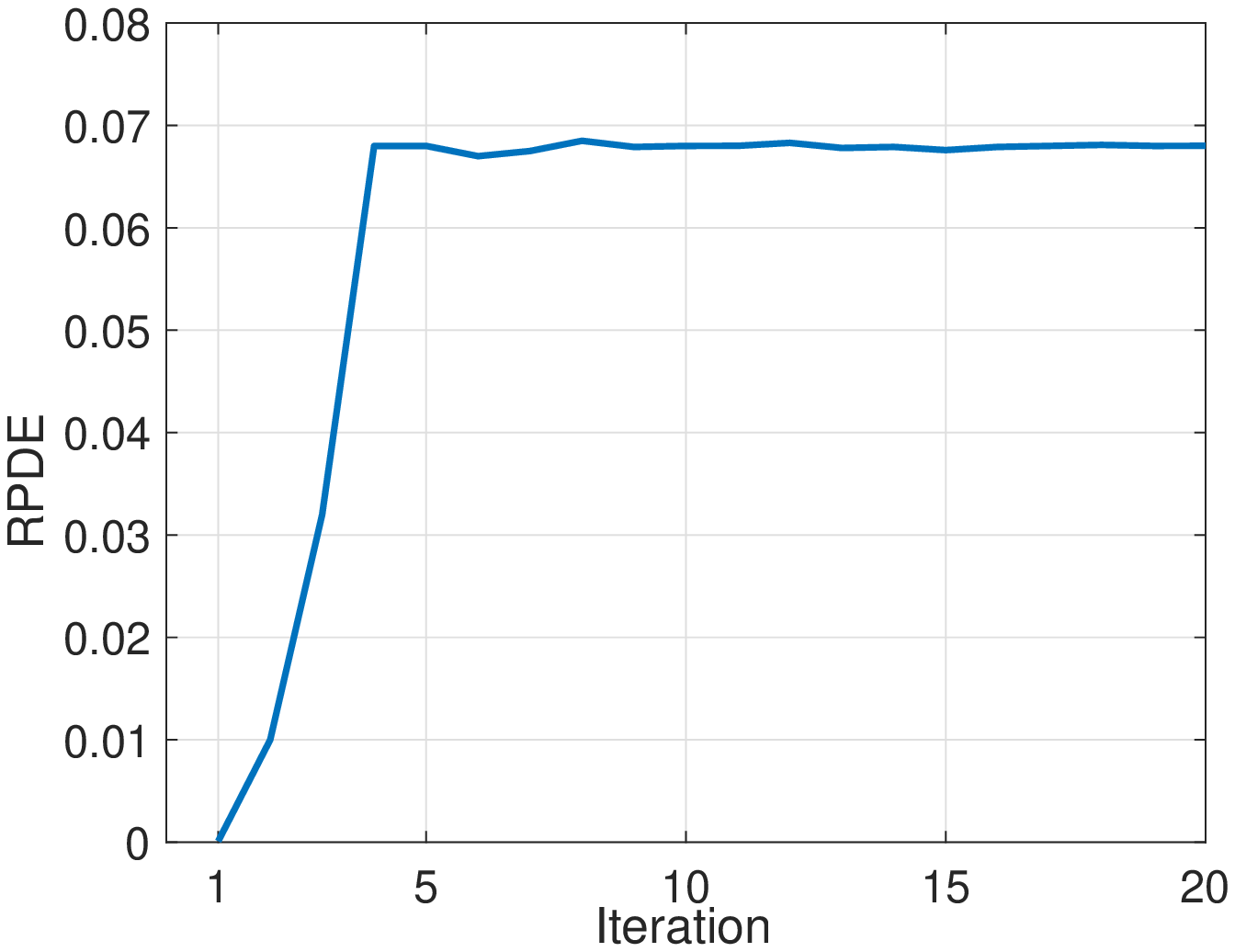}
	\caption{The output RPDE versus the iteration number.}\label{fig:cc}
\end{figure}

In Figure~\ref{fig:bb}, we illustrate the energy distribution processed by the designed algorithm in the two-dimensional range-angle spatial sections.
The angle- and range-dependent outputs provides coherent FDA with the ability to perform joint beamforming in both range and angular dimensions.
As expected, a clear energy peak and extremely low sidelobe appear, indicating that the proposed algorithm can optimally focus the echo energy on the target.
However, for the phased-array or the MIMO radar, due to the lack of the controllable degree of freedom in range dimension, the energy of echo signals can only be steered to a certain angle. 
Figure~\ref{fig:cc} plots the output RPDE versus the number of iterations during the implementation of the algorithm.
It can be observed that our algorithm converges very fast (after about 8-12 iterations).

\subsection{Multiple targets}
\noindent
\begin{figure}[t]
	\centering
	\includegraphics[width=0.5\textwidth]{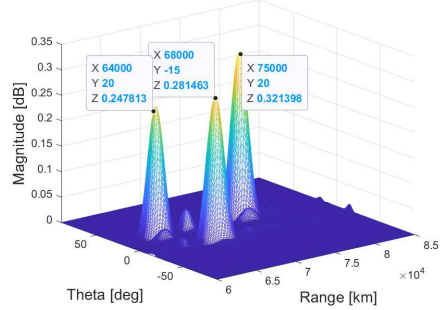}
	\caption{3-D transmit-receive beampattern.}\label{fig:dd}
\end{figure}
Consider the case with three targets, two of which are located at the same angle.
Suppose the potential spatial sections of interest for the transmitter are ${{\Theta }_{d}}=\left[ -{{40}^{o}},-{{10}^{o}} \right]\cup \left[ {{10}^{o}},{{30}^{o}} \right]$ with a mesh grid size $\left( {{{0.1}^o},0.1\kern 1pt km} \right)$.
Three targets are located at $\left( { - {{15}^o},68\kern 1pt km} \right)$, $\left( {{{20}^o},64\kern 1pt km} \right)$, and $\left( {{{20}^o},75\kern 1pt km} \right)$, respectively.
Since the second and third targets have the same angle, traditional radar signal processing based on spatial beamforming cannot separate them.
Figure~\ref{fig:dd} shows the obtained output energy distribution of the coherent FDA.
It can be seen that the received energy is fully concentrated on the assumed target locations, indicating that the FDA joint processing of range and angle can effectively suppress the mainlobe interference at same angle as the target.

\begin{figure}[t]
	\centering
	\includegraphics[width=0.5\textwidth]{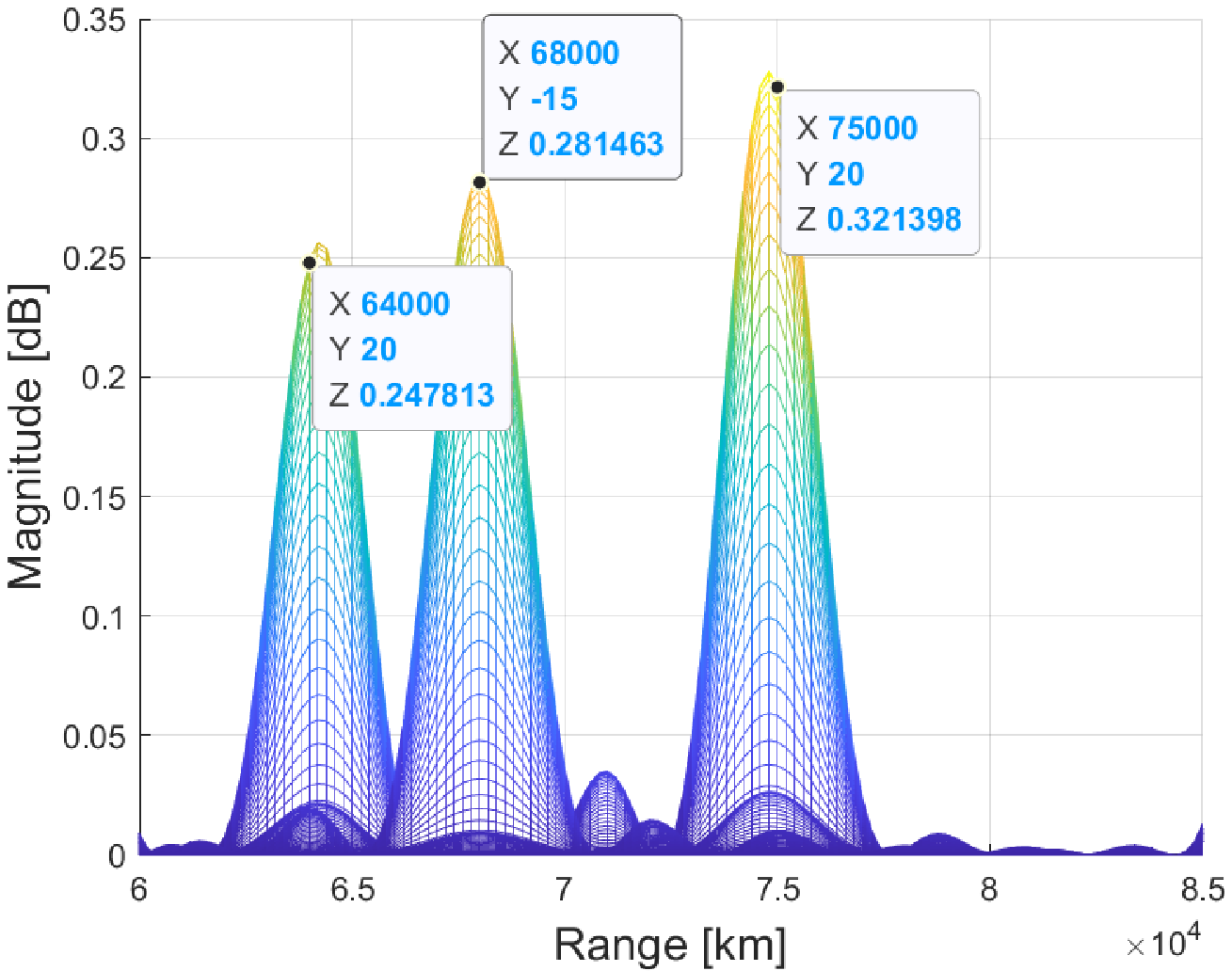}
	\caption{Projection of the transmit-receive beampattern on range dimension.}\label{fig:ee}
\end{figure}

\begin{figure}[t]
	\centering
	\includegraphics[width=0.5\textwidth]{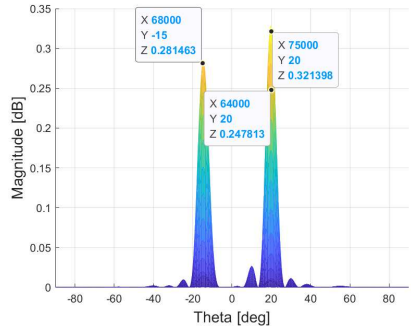}
	\caption{Projection of the transmit-receive beampattern on angle dimension.}\label{fig:ff}
\end{figure}

For clarity, Figures \ref{fig:ee} and \ref{fig:ff} depict the obtained joint transmit-receive beampattern profiles in the range dimension and angle dimension, respectively. 
It is observed that the sidelobes are very low in both range and angle profiles.
It is worth noting that the proposed scheme cannot precisely control the energy allocated to each target. 
Controlling the output of each target greatly increases the complexity of the optimization problem, and this research is left to the future.

\begin{figure}[t]
	\centering
	\includegraphics[width=0.5\textwidth]{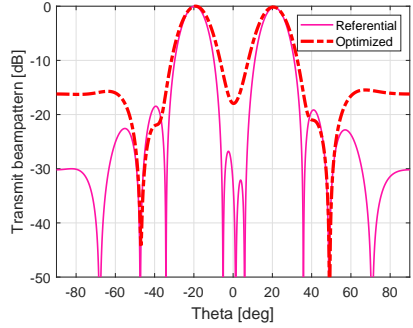}
	\caption{Comparison between the optimized transmit beampattern and the reference transmit beampattern (multiple targets).}\label{fig:gg}
\end{figure}

\begin{figure}[t]
	\centering
	\includegraphics[width=0.5\textwidth]{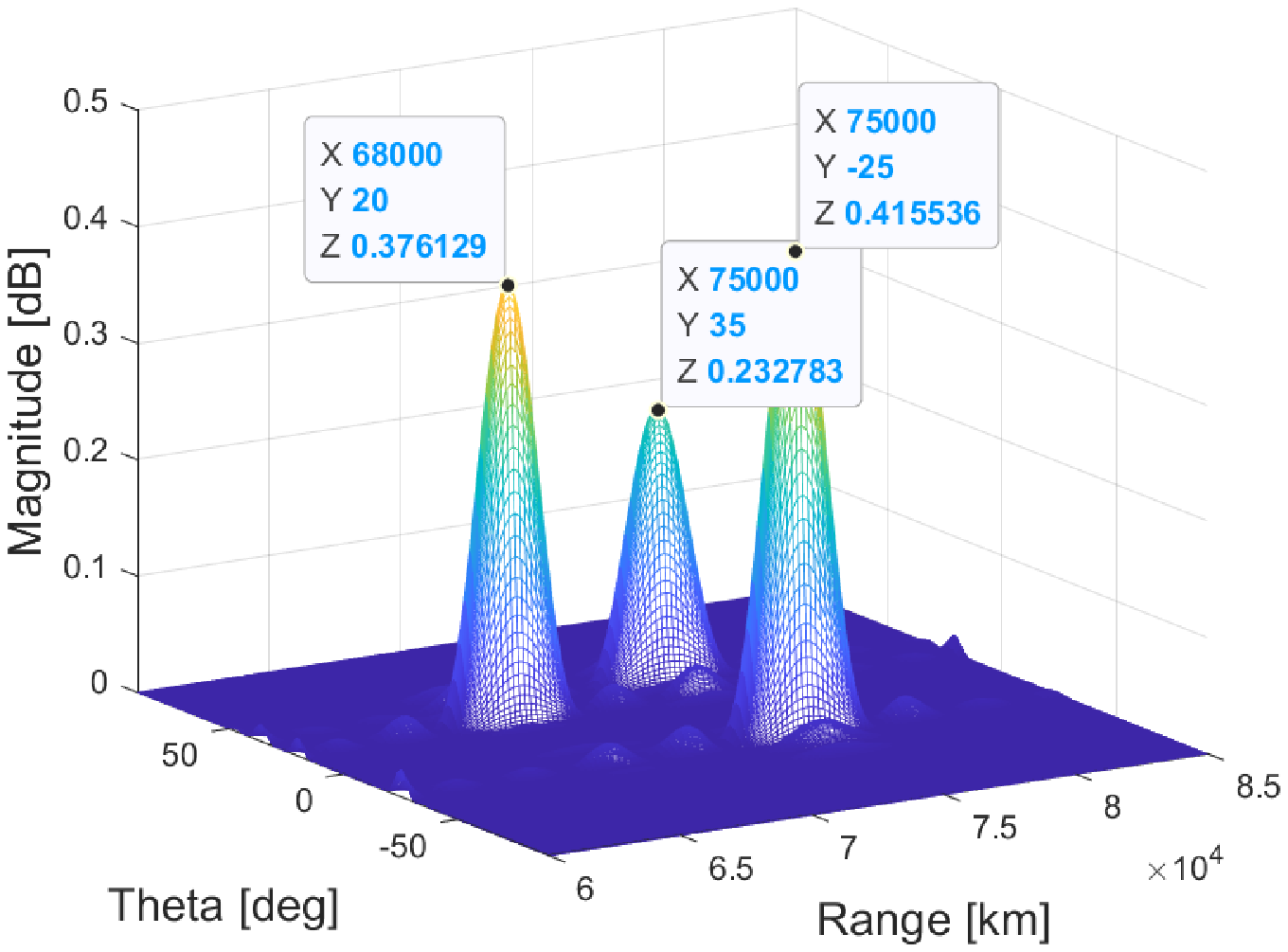}
	\caption{3-D transmit-receive beampattern.}\label{fig:ggg}
\end{figure}

\begin{figure}[t]
	\centering
	\includegraphics[width=0.5\textwidth]{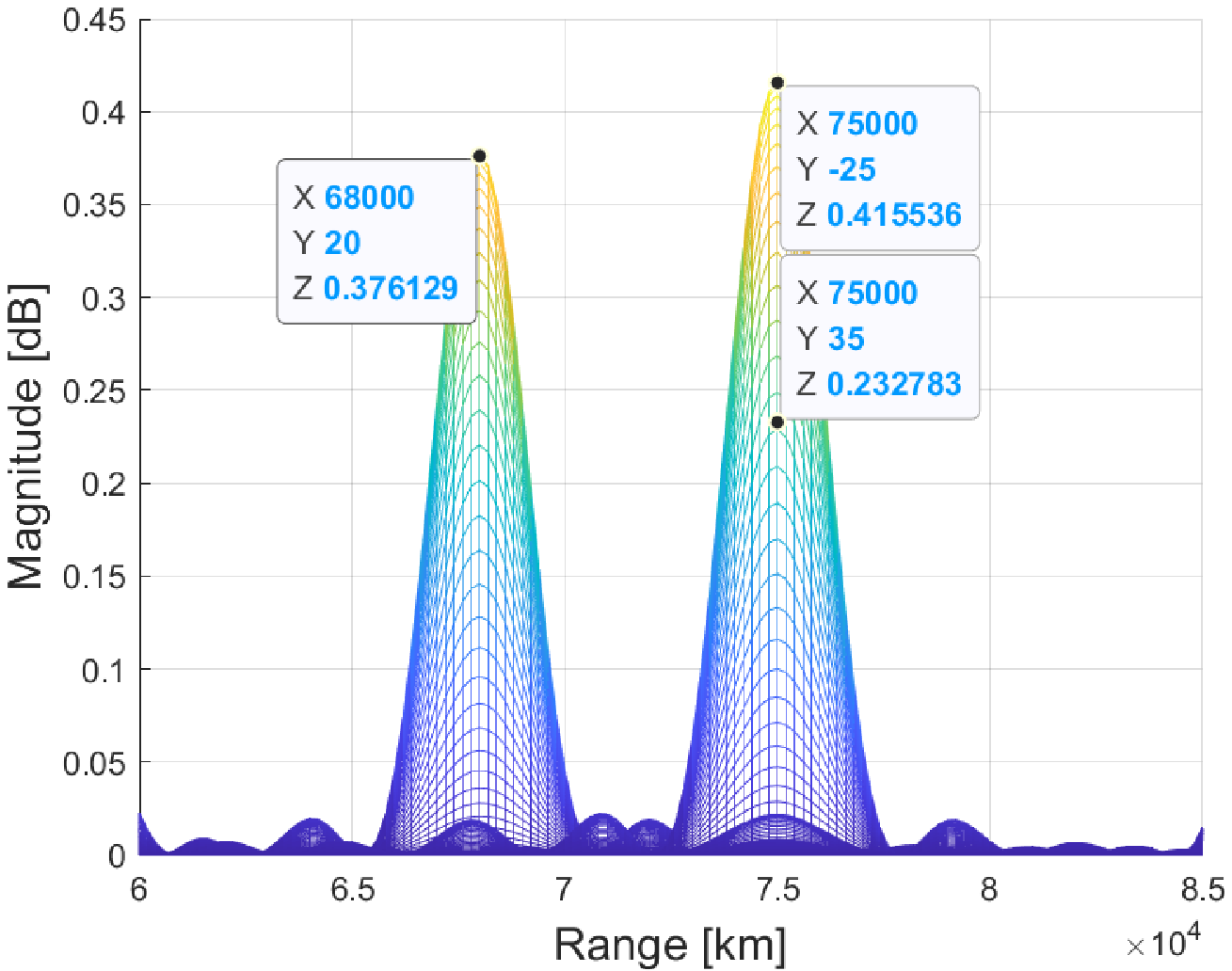}
	\caption{Range profile of the transmit-receive beampattern.}\label{fig:ii}
\end{figure}

\begin{figure}[t]
	\centering
	\includegraphics[width=0.5\textwidth]{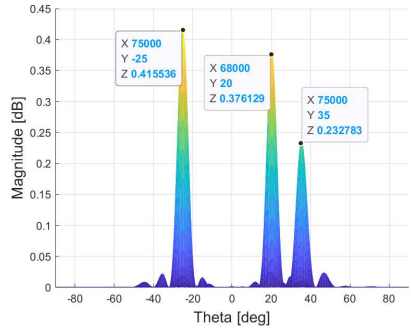}
	\caption{Angle profile of the transmit-receive beampattern.}\label{fig:jj}
\end{figure}

In the second example, we consider another three targets case, two of which are located at the same range. 
The potential spatial sections of interest for transmitter are ${{\Theta }_{d}}=\left[ -{{50}^{o}},-{{10}^{o}} \right]\cup \left[ {{10}^{o}},{{50}^{o}} \right]$ with a mesh grid size $\left( {{{0.1}^o},0.1\kern 1pt km} \right)$.
Three targets are located at $\left( { - {{25}^o},75\kern 1ptkm} \right)$, $\left( {{{20}^o},68\kern 1ptkm} \right)$, and $\left( {{{25}^o},75\kern 1ptkm} \right)$, respectively.
Figure~\ref{fig:gg} compares the optimized transmit beampattern with the reference transmit beampattern.
The widened mainlobe and the increased sidelobe reappear.
In fact, for the proposed transmit-receive scheme, the receive weights mainly control the focusing of echo energy.
For the transmit weights, not only the energy allocation of the transmitting end should be considered, but also the cooperation with the receive weights to achieve the optimal focusing of the echo energy.
Similarly, Figure~\ref{fig:ggg} shows the energy distribution of the coherent FDA.
Figures \ref{fig:ii} and \ref{fig:jj} draw the obtained transmit-receive beampattern profile cuts at the range and angle dimensions, respectively.
It is noted that the coherent FDA can also achieve energy focusing with low sidelobes for different targets in the same range.
This suggests that the coherent FDA has the ability to suppress the range-dependent interferences.

%\begin{figure}[t]
%\vspace{-.5cm}
%\includegraphics[width=3.7in,height = 2.5in]{differenceSPICE_qSPICE} 
%\caption{The resulting estimates of $\tilde\bp$ and $\mathbf{\sigma}$ from the SPICE and the $q$-SPICE estimator ($q$=2). Note that $q$-SPICE is sparser in $\tilde \bp$, whereas SPICE is sparser in $\mathbf{\sigma}$. In this example $r$ is set to $r=1$.} \label{fig:differenceSPICE_qSPICE}
%\end{figure}
\section{Conclusion}\label{sec:e}
\noindent
In this paper, the joint design of transmit and receive weights for coherent FDA radar, considering energy and similarity constraints, has been addressed.
The purpose of this design is to maximize the ratio of the power in the desired two-dimensional range-angle space to the power in the entire area. 
To this end, we formulate the design as an optimization problem, which involves a nonconvex fractional quadratic objective and several quadratic constraints.
In order to tackle the resultant optimization problem, an iterative algorithm based on the SDR and randomization techniques is developed to obtain the optimal solution to this problem. The performance of the devised method is verified by several numerical results. Possible future research might concern the baseband waveform design and optimization of sparse FDA systems for enhanced radar performance. 
\newline

\appendix
\section{}
\label{sec:appen}
Problem $P_1$ can be equivalently recast as the following QCQP optimization problem:
\begin{equation}
\begin{matrix}
\underset{\mathbf{w}}{\mathop{\min }}\, & {{\mathbf{w}}^{H}}\mathbf{\Omega }\left( \Delta f \right)\mathbf{w}  \\
\text{s}\text{.t}\text{.} & {{\mathbf{w}}^{H}}{{\mathbf{\Omega }}_{d}}\left( \Delta f \right)\mathbf{w}=1  \\
\end{matrix}.
\end{equation}
Then, the Lagrange function of above problem can be easily expressed as
\begin{equation}
L\left( {{\mathbf{w}},\upsilon } \right) = {{\mathbf{w}}^H}\mathbf{\Omega }\left( \Delta f \right){\mathbf{w}} + \upsilon  \cdot \left( {{{\mathbf{w}}^H}{{\mathbf{\Omega }}_{d}}\left( \Delta f \right){\mathbf{w}} - 1} \right)
\end{equation}
where $\upsilon$ denotes the Lagrange multiplier.
The Karush-Kuhn-Tucker (KKT) conditions have the form \cite{boyd2004convex}
\begin{equation}
\left\{ \begin{gathered}
\frac{{\partial L\left( {{\mathbf{w}},\upsilon } \right)}}{{\partial {\mathbf{w}}}} = 2\mathbf{\Omega }\left( \Delta f \right){{\mathbf{w}}^ * } + 2\upsilon  \cdot {{\mathbf{\Omega }}_{d}}\left( \Delta f \right){{\mathbf{w}}^ * } = 0 \hfill \\
\upsilon  \geqslant 0 \hfill \\ 
\end{gathered}  \right..
\end{equation}
We have 
\begin{subequations}
	\begin{equation}
	\kern -21pt {{\left( {{\mathbf{w}}^{*}} \right)}^{H}}\mathbf{\Omega }\left( \Delta f \right){{\mathbf{w}}^{*}}+\upsilon \cdot {{\left( {{\mathbf{w}}^{*}} \right)}^{H}}{{\mathbf{\Omega }}_{d}}\left( \Delta f \right){{\mathbf{w}}^{*}}=0
	\end{equation}
	\begin{equation}
	\kern 77pt {{\mathbf{\Omega }}^{-1}}\left( \Delta f \right){{\mathbf{\Omega }}_{d}}\left( \Delta f \right){{\mathbf{w}}^{*}}=-\frac{1}{\upsilon }\cdot {{\mathbf{w}}^{*}}.
	\end{equation}
\end{subequations}
One may conclude that
\begin{itemize}
	\item[1)] Minimizing ${{\mathbf{w}}^H}\mathbf{\Omega }\left( \Delta f \right){\mathbf{w}}$ is equivalent to minimizing $- \upsilon $;
	\item[2)] ${{\mathbf{w}}^ * }$ is an eigenvector of matrix ${{\mathbf{\Omega }}^{-1}}\left( \Delta f \right){{\mathbf{\Omega }}_{d}}\left( \Delta f \right)$, and the corresponding eigenvalue is $ - \frac{1}{\upsilon }$.
\end{itemize}
Therefore, we can infer that the optimal reference weight vector ${{\mathbf{w}}_0 }$ is the eigenvector corresponding to the maximum eigenvalue of matrix ${{\mathbf{\Omega }}^{-1}}\left( \Delta f \right){{\mathbf{\Omega }}_{d}}\left( \Delta f \right)$.
This completes the proof.

\bibliographystyle{unsrt}
\bibliography{Refs}

\end{document}